\documentclass[12pt]{article}
\usepackage{graphicx}
\def\hybrid{\topmargin 0pt      \oddsidemargin 0pt
        \headheight 0pt \headsep 0pt
        \voffset=-0.5cm
        \textwidth 6.25in       
        \textheight 9.5in       
        \marginparwidth 0.0in
        \parskip 5pt plus 1pt   \jot = 1.5ex}
\catcode`\@=11
\def\marginnote#1{}

\newcount\hour
\newcount\minute
\newtoks\amorpm
\hour=\time\divide\hour by60
\minute=\time{\multiply\hour by60 \global\advance\minute by-\hour}
\edef\standardtime{{\ifnum\hour<12 \global\amorpm={am}%
        \else\global\amorpm={pm}\advance\hour by-12 \fi
        \ifnum\hour=0 \hour=12 \fi
        \number\hour:\ifnum\minute<10 0\fi\number\minute\the\amorpm}}
\edef\militarytime{\number\hour:\ifnum\minute<10 0\fi\number\minute}

\def\draftlabel#1{{\@bsphack\if@filesw {\let\thepage\relax
   \xdef\@gtempa{\write\@auxout{\string
      \newlabel{#1}{{\@currentlabel}{\thepage}}}}}\@gtempa
   \if@nobreak \ifvmode\nobreak\fi\fi\fi\@esphack}
        \gdef\@eqnlabel{#1}}
\def\@eqnlabel{}
\def\@vacuum{}
\def\draftmarginnote#1{\marginpar{\raggedright\scriptsize\tt#1}}
\def\draftlabel#1{{\@bsphack\if@filesw {\let\thepage\relax
   \xdef\@gtempa{\write\@auxout{\string
      \newlabel{#1}{{\@currentlabel}{\thepage}}}}}\@gtempa
   \if@nobreak \ifvmode\nobreak\fi\fi\fi\@esphack}
        \gdef\@eqnlabel{#1}}
\def\@eqnlabel{}
\def\@vacuum{}
\def\draftmarginnote#1{\marginpar{\raggedright\scriptsize\tt#1}}

\def\draft{\oddsidemargin -.5truein
        \def\@oddfoot{\sl preliminary draft \hfil
        \rm\thepage\hfil\sl\today\quad\militarytime}
        \let\@evenfoot\@oddfoot \overfullrule 3pt
        \let\label=\draftlabel
        \let\marginnote=\draftmarginnote
   \def\@eqnnum{(\theequation)\rlap{\kern\marginparsep\tt\@eqnlabel}%
\global\let\@eqnlabel\@vacuum}  }

\def\numberbysection{\@addtoreset{equation}{section}
        \def\theequation{\thesection.\arabic{equation}}}

\def\underline#1{\relax\ifmmode\@@underline#1\else
        $\@@underline{\hbox{#1}}$\relax\fi}

\def\titlepage{\@restonecolfalse\if@twocolumn\@restonecoltrue\onecolumn
     \else \newpage \fi \thispagestyle{empty}\c@page\z@
        \def\thefootnote{\fnsymbol{footnote}} }

\def\endtitlepage{\if@restonecol\twocolumn \else  \fi
        \def\thefootnote{\arabic{footnote}}
        \setcounter{footnote}{0}}  
\relax

\numberbysection
\hybrid

\newfont{\Bbb}{msbm10 scaled 1\@ptsize00}
\newfont{\Bbbb}{msbm7 scaled 1\@ptsize00}
\newcommand{\CC}{\mbox{\Bbb C}}
\newcommand{\CCC}{\mbox{\Bbbb C}}
\newcommand{\DD}{\mbox{\Bbb D}}
\newcommand{\DDD}{\raise-1pt\hbox{$\mbox{\Bbbb D}$}}
\newcommand{\HH}{\mbox{\Bbb H}}
\newcommand{\HHH}{\mbox{\Bbbb H}}

\newcommand{\UUU}{\raise-1pt\hbox{$\mbox{\Bbbb U}$}}

\newcommand{\ZZ}{\mbox{\Bbb Z}}
\newcommand{\z}{\raise-1pt\hbox{$\mbox{\Bbbb Z}$}}

\def\beq{\begin{equation}}
\def\eeq{\end{equation}}
\def\p{\partial}

\def\psistar{\psi^{*}}

\def\normord{ {\scriptstyle {{\bullet}\atop{\bullet}}} }

\def\normordbare{ {\scriptstyle {{ \times}\atop{ \times}}} }

\def\rbr{\right >}

\def\lvac{\left <0\right |}
\def\rvac{\left |0\right >}

\def\lvacn{\left <n\right |}
\def\rvacn{\left |n\right >}
\def\lvacN{\left <N\right |}
\def\rvacN{\left |N\right >}

\begin{document}
\begin{titlepage}

\title{Canonical and grand canonical
partition functions of Dyson gases as tau-functions
of integrable hierarchies
and their fermionic realization}

\author{A.~Zabrodin
\thanks{Institute of Biochemical Physics,
4 Kosygina st., 119334, Moscow, Russia and ITEP, 25
B.Cheremushkinskaya, 117218, Moscow, Russia}}

\date{January 2010}
\maketitle

\begin{abstract}

The partition function for a canonical ensemble of 2D Coulomb
charges in a background potential (the Dyson gas) is
realized as a vacuum expectation value of a
group-like element constructed in terms of free fermionic
operators. This representation provides an explicit identification
of the partition function with a tau-function of the
2D Toda lattice hierarchy.
Its dispersionless (quasiclassical) limit yields
the tau-function for analytic curves encoding the
integrable structure of the inverse
potential problem and parametric conformal maps.
A similar fermionic realization of partition
functions for grand canonical ensembles of
2D Coulomb charges in the presence of
an ideal conductor is also suggested. Their
representation as Fredholm determinants is given
and their
relation to integrable hierarchies, growth problems and
conformal maps is discussed.

\end{abstract}
\vfill

\end{titlepage}

\section{Introduction}

Statistical ensembles of 2D Coulomb particles (referred to also as
logarithmic gases, $\beta$-ensembles or Dyson gases) at a particular
value of inverse temperature, $\beta =2$, are known to have
remarkable integrable properties. Their partition functions,
regarded as functions of properly chosen parameters of the
background trapping potential (coupling constants), can be
identified with tau-functions for hierarchies of nonlinear
integrable equations such as 2D Toda lattice (2DTL) or
Kadomtsev-Petviashvili (KP) hierarchies. Equivalently, the
tau-function provides a generating series for correlation functions
of the Dyson gas.

The aim of this paper is to give an explicit representation of
the partition functions as vacuum expectation values of certain
operators constructed from free fermions, in the spirit of
the Kyoto school \cite{DJKM83,JM83}. Similar constructions
in the context of random matrix models of different types
were given in \cite{KMMOZ91,KMMM93,HO03,HO06}.
However, most of the previous studies were devoted to canonical
ensembles with a fixed number of particles, $N$, which after
F. Dyson \cite{Dyson} are customarily viewed as
eigenvalues of a $N\times N$ random matrix. The canonical
partition function, $Z_N$, is then
tau-function of the 2DTL hierarchy with the discrete variable
$N$.

A natural question is whether this construction can be
extended to grand canonical ensembles.
Taking a weighted sum of the $Z_N$'s,
${\cal Z}=\sum_{N}e^{\mu N}Z_N$, one obtains
a quantity whose interpretation in terms of integrable hierarchies
is presently not clear.
A way to introduce a grand canonical ensemble of
2D Coulomb charges with transparent integrable
properties was first suggested in \cite{LS00}.
The idea is to consider the logarithmic gas with varying
number of particles in the presence
of an ideal conductor, so that each particle interacts not only
with other particles of the gas but also with their ``mirror
images" of opposite charge.
The grand canonical partition functions
of such systems appear to be tau-functions of the 2DTL or KP
hierarchies (depending on whether the conductor fills a disk or
a half-plane), whose ``times" again serve as parameters of the
background potential but in a different manner. The fermionic
operator construction of these tau-functions
is technically even simpler than that for canonical ensembles.

Our motivation comes from growth problems
of Laplacian type such as
viscous flows in the Hele-Show cell \cite{list,book}.
At zero surface tension, the model has an integrable structure
of the 2DTL hierarchy in the zero dispersion limit \cite{MWWZ00}.
From mathematical point of view, the same integrable hierarchy
stays behind some classical problems of complex analysis:
parametric deformations of conformal maps \cite{WZ00},
boundary value problems for Laplace operator \cite{MWZ02,KMZ05} and
the inverse potential problem in two dimensions \cite{Z01}.
Switching on the dispersion thus coming back to the
original 2DTL hierarchy, one obtains a non-trivial
deformation of this integrable structure. However,
the corresponding deformation of Laplacian growth
and, what would be even more intriguing,
of the above mentioned problems of complex analysis,
is not yet formulated explicitly in proper terms.

The theory of logarithmic gases provides another
view on these matters \cite{Z03,TBAZW05}.
In the thermodynamic limit,
when the number of particles is very large,
the logarithmic gas in the leading approximation
macroscopically looks like a charged fluid
with continuous density. The equilibrium state of the
system is a result of competition between the mutual repelling
of particles, which tends to remove them to infinity, and
the external force which attracts each particle to local
minima of the background potential.
Typically, in the equilibrium the fluid occupies
some compact domains in the plane around local
minima of the background potential, which we call droplets.
The shape of the droplets
is determined by their total charge and by
the profile of the trapping potential.
When one increases the total charge keeping the potential
fixed, the droplet changes its shape according to the
growth law specific for the Laplacian
growth processes (the Darcy's law).
At the same time, the integrable deformation
of the $N=\infty$ picture
is naturally build in the 2D Coulomb gas.
We believe that the reformulation of the model
in terms of free fermions may help to clarify the geometrical
meaning of this deformation.

Another interesting question is
what kind of growth problems emerge in the thermodynamic
limit of the grand canonical ensembles. We give a partial
answer restricting ourselves to the grand canonical ensemble
in the upper half-plane, in which case the growth problem
is equivalent to Laplacian growth of ``fat slits" considered
in \cite{Z09}.

\section{Free fermions and tau-functions}

In this section we recall the free
fermionic construction \cite{DJKM83,JM83}
of KP and 2DTL tau-functions.

Let $\psi_n , \psistar_{n}$, $n\in \ZZ$, be free fermionic
operators with usual anticommutation relations
$[\psi_n , \psi_m ]_+ = [\psistar_n, \psistar_m]_+=0$,
$[\psi_n , \psistar_m]_+=\delta_{mn}$. They generate
an infinite dimensional Clifford algebra. We also use their
Fourier transforms
\beq\label{ferm0}
\psi (z)=\sum_{k\in \z}\psi_k z^k, \quad \quad
\psistar (z)=\sum_{k\in \z}\psistar_k z^{-k}
\eeq
which are regarded as free fermionic fields in the
complex plane of the variable $z$.

Neutral bilinear combinations
$\sum A_{mn}\psi_m \psistar_n$
of the fermions generate the Lie algebra
$gl(\infty )$. Normal ordering of the generators
(see below) allows one
to consider certain infinite sums as well.
Exponentiating these expressions, one obtains
an infinite dimensional group
which is a central extension of $GL_{\infty}$.
Let us call elements
of the Clifford algebra of the form
$g=\exp \left (\sum_{mn}A_{mn}\psi_m \psistar_n\right )$ group-like
elements.
A characteristic property of the group-like elements $g$ is that
$g\psi _n g^{-1}$ is a linear combination of $\psi_j$'s
and similarly for $\psistar_n$: $g\psi _n g^{-1} = \sum_l \psi_l R_{ln}$,
$g\psistar _n g^{-1} = \sum_l \psistar_l R^{-1}_{nl}$, where the matrix
$R$ is determined by the matrix $A$. Of particular
importance are the group-like elements obtained by
exponentiating the operators
\beq\label{ferm2}
J_+ = \sum_{k\geq 1}t_k J_k, \quad \quad
J_- = \sum_{k\geq 1}t_{-k} J_{-k}
\eeq
where
$$
J_k =\sum_{j\in \z}\psi_j \psistar_{j+k}
$$
are Fourier modes of the ``current operator"
and $t_k$ are parameters (called times). In general,
they are complex numbers.
It is convenient to denote the collection of times
with positive (negative) indices
by $t_+ = \{t_1, t_2 , \ldots \}$ and $t_- = \{t_{-1}, t_{-2} , \ldots \}$
respectively and to introduce their generating function
$$
\xi (t_{\pm}, z)=\sum_{k\geq 1}t_{\pm k}z^k.
$$
It is easy to check that the fields $\psi (z)$, $\psistar (z)$
transform diagonally under the adjoint action
of the group-like elements
$e^{J_+}$, $e^{J_-}$:
\beq\label{ferm3}
\begin{array}{l}
e^{J_{\pm}}\psi (z)e^{-J_{\pm}}=e^{\xi (t_{\pm} ,\, z^{\pm 1})}\psi (z)
\\ \\
e^{J_{\pm}}\psistar (z)e^{-J_{\pm}}=e^{-\xi (t_{\pm} , \,
z^{\pm 1})}\psistar (z).
\end{array}
\eeq
In terms of polynomials $p_k(t_{\pm})$ defined by
\beq\label{schur1}
e^{\xi (t_{\pm} ,\, z)}=\sum_{k\geq 0}p_k (t_{\pm})z^k
\eeq
the corresponding formulas for $\psi_n , \psistar_{n}$
can be written as
\beq\label{ferm4}
\begin{array}{l}
\displaystyle{
e^{J_{\pm}}\psi _n e^{-J_{\pm}}=
\sum_{k\geq 0}\psi_{n\mp k}p_k(t_{\pm})}
\\ \\
\displaystyle{
e^{J_{\pm}}\psistar _n e^{-J_{\pm}}=
\sum_{k\geq 0}\psistar_{n\pm k}p_k(-t_{\pm})}.
\end{array}
\eeq

\begin{figure}[t]
   \centering
        \includegraphics[angle=-00,scale=0.45]{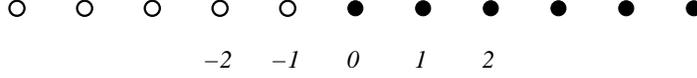}
        \caption{\it The Dirac sea $\rvac$. Filled states are shown in
        black, empty ones in white.}
    \label{fig:vac}
\end{figure}

Next, we introduce a vacuum state $\left |0\rbr$ which is
a ``Dirac sea" where all negative mode states are empty
and all positive ones are occupied (Fig. \ref{fig:vac}):
$$
\psi_n \rvac =0, \quad n< 0; \quad \quad \quad
\psistar_n \rvac =0, \quad n\geq 0.
$$
(For brevity, we call indices $n\geq 0$ {\it positive}.)
With respect to this vacuum, the operators $\psi_n$ with
$n<0$ and $\psistar_n$ with $n\geq 0$ are annihilation operators
while the operators $\psistar_n$ with $n<0$ and
$\psi_n$ with $n\geq 0$ are creation operators.
Similarly, the dual vacuum state has the properties
$$
\lvac \psistar_n  =0, \quad n< 0; \quad \quad \quad
\lvac \psi_n  =0, \quad n\geq 0.
$$
We also need ``shifted" Dirac vacua $\rvacn$ and $\lvacn$
defined as
$$
\rvacn = \left \{
\begin{array}{l}
\psi_{n-1}\ldots \psi_1 \psi_0 \rvac , \,\,\,\,\, n> 0
\\ \\
\psistar_n \ldots \psistar_{-2}\psistar_{-1}\rvac , \,\,\,\,\, n<0
\end{array} \right.
$$
$$
\lvacn = \left \{
\begin{array}{l}
\lvac \psistar_{0}\psistar_{1}\ldots \psistar_{n-1} , \,\,\,\,\, n> 0
\\ \\
\lvac \psi_{-1}\psi_{-2}\ldots \psi_{n} , \,\,\,\,\, n<0
\end{array} \right.
$$

The vacuum expectation value $\lvac \ldots \rvac$ is a
hermitian linear form on the Clifford algebra defined
on bilinear combinations of fermions
by the properties $\left. \lvac \! 0\right > =1$,
$\lvac \psi_n \psi_m\rvac = \lvac \psistar_n \psistar_m \rvac =0$
for all $m,n$ and
$$
\lvac \psi_n \psistar_m\rvac =\delta_{mn}\quad
\mbox{for $m<0$}, \quad \quad
\lvac \psi_n \psistar_m\rvac =0 \quad \mbox{for $m\geq 0$}.
$$
Its extension to the whole algebra is given by the Wick's theorem:
Let $w_i$ be arbitrary linear combinations of $\psi_j$ and $\psistar_j$,
then $\lvac w_1\ldots w_{2n+1}\rvac =0$ and
$$
\lvac w_1\ldots w_{2n}\rvac =\sum_{\sigma} (-1)^{P(\sigma )}
\lvac w_{\sigma (1)} w_{\sigma (2)}\rvac \ldots
\lvac w_{\sigma (2n-1)} w_{\sigma (2n)}\rvac .
$$
Here the sum is over permutations $\sigma$ of $1, \ldots , 2n$
such that $\sigma (1)<\sigma (2)$, $\sigma (3)<\sigma (4)$, $\ldots$,
$\sigma (2n-1)<\sigma (2n)$ and $\sigma (1)<\sigma (3)< \ldots <
\sigma (2n-1)$ and
$P(\sigma )$ is the parity of $\sigma$.
The Wick's theorem is often used in the following form.
Let $w_i$ be linear combinations of $\psi_j$'s only and
$w_i^*$ be linear combinations of $\psistar_j$'s only, then
$$
\lvac w_1 \ldots w_n w_n^* \ldots w_1^* \rvac =
\det_{i,j =1,\ldots , n}\lvac w_i w_j^* \rvac .
$$
Let us also give an explicit formula for the expectation
value of products of the fields $\psi (z)$, $\psistar (\zeta )$:
\beq\label{ferm5}
\begin{array}{c}
\displaystyle{
\lvacN \psi (z_1)\ldots \psi (z_n)
\psistar (\zeta _n)\ldots \psistar (\zeta _1)\rvacN =
\prod_{l=1}^{n} (z_l / \zeta_l)^N \cdot \, \det_{i,j}
\frac{\zeta _i}{z_i-\zeta_j}}
\\  \\
\displaystyle{
\phantom{aaaaaaaaaaaaaaa}
=\,\, \prod_{l} \frac{z_{l}^{N}\zeta_{l}^{1-N}}{(z_l -\zeta_l)}\,\,
\prod_{i<j}\frac{(z_i -z_j)(\zeta_j -\zeta_i)}{(z_i-\zeta_j)(z_j-\zeta_i)}}.
\end{array}
\eeq

One may define the normal ordering $\normord (\ldots )\normord $ with respect
to the Dirac vacuum $\rvac$ when all annihilation operators
are moved to the right and all creation operators are moved to
the left taking into account their anticommutativity under mutual
permutations. For example,
$\normord  \psistar_{1}\psi_{1}\normord =
-\psi_{1} \psistar_{1}=\psistar_{1}\psi_{1} -1$,
and, more generally,
$\normord \psistar_m \psi_n\normord =
\psistar_m \psi_n -\lvac \psistar_m \psi_n \rvac$.
We also note the identities
\beq\label{ferm7}
\begin{array}{l}
e^{\alpha \psi_k \psistar_k}=1+(e^{\alpha}-1)\psi_k \psistar_k =
\normord e^{(e^{\alpha}-1)\psi_k \psistar_k}\normord \quad \quad
\mbox{for $k\geq 0$},
\\ \\
e^{\alpha \psistar_k \psi_k}=1+(e^{\alpha}-1) \psistar_k \psi_k =
\normord e^{(e^{\alpha}-1) \psistar_k \psi_k}\normord \quad \quad
\mbox{for $k< 0$}.
\end{array}
\eeq

As it has been established in the works of the Kyoto school,
the expectation values of group-like elements are tau-functions
of integrable hierarchies of nonlinear differential equations.
In particular,
\beq\label{ferm6}
\tau_n (t_+ , t_-)=\lvacn e^{J_+}ge^{-J_-}\rvacn
\eeq
is the tau-function of the 2DTL hierarchy
meaning that it obeys the infinite set of Hirota bilinear equations
for the 2DTL hierarchy the simplest of which is
$$
\p_{t_1}\! \tau_n \, \p_{t_{-1}}\! \tau_n - \tau_n \,
\p_{t_1}\p_{t_{-1}}\! \tau_n
=\tau_{n+1}\tau_{n-1}.
$$
In a similar manner,
\beq\label{ferm6a}
\tau (t_+)=\lvac e^{J_+}g\rvac
\eeq
is the tau-function of the KP hierarchy.

It is worthwhile to note that the group-like elements of the
Clifford algebra
can be also written as normal ordered exponents of the form
\beq\label{ferm1}
g=\normord \exp \left (\sum_{mn}\tilde A_{mn}\psi_m \psistar_n\right )
\normord
\eeq
For regular (invertible) elements $g$, this normal ordered
expression is just equal, modulo a constant factor, to an exponent of a
fermionic bilinear form without normal ordering
but with some other matrix $\tilde A$ (an example is provided
by (\ref{ferm7}), see also \cite{SMJ} for a more general
case). If $g$ is invertible, then the
tau-function at any $n$ is not identically zero.
However, there is an important class of
tau-functions which vanish identically at some values of $n$.
They correspond to {\it singular} elements $g$ which are not invertible
(and thus do not belong to a group in a strict sense) but still
can be represented in the normal ordered form (\ref{ferm1}).

Here are two important examples of such singular elements:
\beq\label{ferm8}
\begin{array}{l}
\displaystyle{
P_{+}= \, \normord \exp \left (\sum_{i<0}\psi_i \psistar_i\right )\normord }
=\, \prod_{i<0}(1-\psistar_i \psi_i )=\, \prod_{i<0}\psi_i \psistar_i ,
\\ \\
\displaystyle{
P_{-}= \normord \exp \left (-\! \sum_{i\geq 0}\psi_i \psistar_i\right )\normord }
=\prod_{i\geq 0}(1-\psi_i \psistar_i )=\prod_{i\geq 0}\psistar_i \psi_i .
\end{array}
\eeq
In a sense, these operators are projectors to positive and
negative modes respectively. Their properties (extensively
used in what follows)
can be easily seen from the definition. The both operators
obey the projector property:
$P_{\pm}^2 =P_{\pm}$. The operator $P_+$ kills negative
creation modes standing to the right and negative annihilation modes
standing to the left and commutes with all positive modes:
\beq\label{ferm9}
\begin{array}{l}
\phantom{a}P_+ \psistar_k =\psi_k P_+ =0, \quad \quad \quad k<0,
\\ \\ \phantom{a} [P_+ , \psistar_k ]=[P_+ , \psi_k ]=0, \quad k \geq 0.
\end{array}
\eeq
The operator $P_-$ kills positive
creation modes standing to the right and positive annihilation modes
standing to the left and commutes with all negative modes:
\beq\label{ferm9a}
\begin{array}{l}
\phantom{a}P_- \psi_k =\psistar_k P_- =0, \quad \quad \quad k\geq 0,
\\ \\ \phantom{a} [P_- , \psistar_k ]=[P_+ , \psi_k ]=0, \quad k < 0.
\end{array}
\eeq
From this it is obvious that $P_+ \! \rvacn =0$ at $n<0$ and
$P_+ \! \rvacn =\rvacn$ at $n\geq 0$. Similarly,
$P_- \! \rvacn =0$ at $n\geq 0$ and
$P_- \! \rvacn =\rvacn$ at $n<0$. Somewhat less obvious properties
(also used in what follows) are $P_+ e^{-J_-}\rvac =\rvac$,
$\lvac e^{J_+}P_+ =\lvac$.

As an example, let us calculate the tau-function corresponding
to the singular element $P_+$:
$\tau_N =\lvacN e^{J_+} P_+ e^{-J_-}\rvacN$.
Here we follow \cite{KMMOZ91}. First of all, it is
not difficult to see that
$\tau_N =0$ at $N<0$ and $\tau_0 =1$. For $N>1$ we have:
$$
\tau_N = \lvac \psistar_{N-1}\ldots \psistar_0 \, e^{J_+}P_+
e^{-J_-}\psi_0 \ldots \psi_{N-1} \rvac .
$$
To proceed, it is convenient to use the short hand notation
$\psi_n (H)=e^H \psi_n e^{-H}$, $\psistar_n (H)=e^H \psistar_n e^{-H}$,
where $H$ is any operator from the Clifford algebra, then we can write
$$
\tau_N = \lvac e^{J_+} \psistar_{N-1}(-J_+)\ldots \psistar_0 (-J_+) P_+
P_+ \psi_0 (-J_-)\ldots \psi_{N-1}(-J_-) e^{-J_-} \rvac .
$$
Equations (\ref{ferm4}) imply
\beq\label{ferm4a}
\begin{array}{l}
\displaystyle{
\psi _n (-J_-)=
\sum_{k\geq 0}\psi_{n+k}\, p_k(-t_{-})}
\\ \\
\displaystyle{
\psistar _n (-J_+)=
\sum_{k\geq 0}\psistar_{n+k} \, p_k(t_{+})}.
\end{array}
\eeq
so the operators $\psi _n (-J_-)$, $\psistar _n (-J_+)$ in the
formula for $\tau_N$ contain only positive modes and, therefore,
commute with $P_+$. Moving one $P_+$ to the right and another
one to the left, and using the properties mentioned above,
we obtain
$$
\begin{array}{lll}
\tau_N &=&\displaystyle{
\lvac  \psistar_{N-1}(-J_+)\ldots \psistar_0 (-J_+)\,
\psi_0 (-J_-)\ldots \psi_{N-1}(-J_-)  \rvac }
\\ && \\
&=&\displaystyle{
\det_{1\leq j,k \leq N}\lvac \psistar_{j-1}(-J_+)\psi_{k-1} (-J_-)\rvac }
\end{array}
$$
by the Wick's theorem. The expectation value under the determinant
can be represented as a contour integral as follows:
$$
\begin{array}{c}
\displaystyle{
\lvac \psistar_{j}(-J_+)\psi_{k} (-J_-)\rvac =
\sum_{a,b\geq 0}p_a (t_+)p_b (-t_-)\lvac \psistar_{j+a}\psi_{k+b}\rvac}
\\ \\
\displaystyle{ =\,
\sum_{a,b\geq 0}p_a (t_+)p_b (-t_-)\delta_{j+a, k+b}=
\oint_{|z|=1} z^{j-k} e^{\xi (t_+, z)-\xi (t_- , 1/z)}
\frac{dz}{2\pi i z}}\, .
\end{array}
$$
The whole determinant can then be written as an $N$-fold
contour integral:
\beq\label{unitary}
\tau_N = \frac{1}{N!}\oint \ldots \oint
\prod_{j<k}(z_j -z_k)(z_{j}^{-1}-z_{k}^{-1})
\prod_{l=1}^{N}e^{\xi (t_+, z_l)-\xi (t_- , 1/z_l)}\frac{dz_l}{2\pi i z_l}\, .
\eeq
When $t_{-k}=-\bar t_k$,
the expression $\xi (t_+, z)-\xi (t_- , 1/z)$ is purely real
for $z$ on the unit circle and
$\tau_N$ coincides with the partition function of the unitary
random matrix model written in terms of the eigenvalues.
In this form, it can be treated also as the partition function
of a canonical ensemble of $N$ 2D Coulomb particles confined
on a circle.

\section{Partition function of the 2D Coulomb gas as a
tau-function: canonical ensemble}

Let us fix an arbitrary measure $d\mu (z)$ in the complex plane
and consider the following group-like element:
\beq\label{can1}
g_0 = \normord \exp \left (\int_{\CCC}\psi_+(z)\psistar _{+}(1/\bar z)
d\mu (z) -\sum_{j\geq 0} \psi_j \psistar_j \right )\normord
\eeq
Here $\displaystyle{\psi_+(z)= \sum_{n\geq 0}\psi_n z^n}$,
$\displaystyle{\psistar_+(z)= \sum_{n\geq 0}\psistar_n z^{-n}}$ are
truncated Fourier series containing only positive modes. Obviously,
$g_0$ commutes with $P_+$. Expending the exponent into a series,
one can represent $g_0$ in a more explicit form:
\beq\label{can2}
g_0= \sum_{m=0}^{\infty} \frac{1}{m!}
\int_{\CCC ^m} \psi_+ (z_1) \ldots \psi_+(z_m)P_-
\psistar_+ (1/\bar z_m)\ldots \psistar_+ (1/\bar z_1)
d\mu_1 \ldots d\mu_m\,,
\eeq
where $d\mu_j \equiv d\mu (z_j)$.

Let us consider the expectation value
\beq\label{can3}
\tau_N (t_+,t_-)=\lvacN e^{J_+}g_0 P_+ e^{-J_-}\rvacN
\eeq
and apply to it a chain of transformations similar to the ones
used in the simpler case $g_0=1$. Again, $\tau_N =0$ at $N<0$ and $\tau_0 =1$.
For $N>1$ we have:
$$
\begin{array}{lll}
\tau_N &=& \lvac \psistar_{N-1}\ldots \psistar_0 \, e^{J_+}  g_0 P_+
e^{-J_-}\psi_0 \ldots \psi_{N-1} \rvac
\\ && \\
&=&\lvac e^{J_+} \psistar_{N-1}(-J_+) \ldots \psistar_0 (-J_+) P_+  g_0 P_+
\psi_0 (-J_-)\ldots \psi_{N-1}(-J_-) e^{-J_-} \rvac
\\ && \\
&=&\lvac  \psistar_{N-1}(-J_+) \ldots \psistar_0 (-J_+)   g_0
\psi_0 (-J_-)\ldots \psi_{N-1}(-J_-)  \rvac .
\end{array}
$$
Substituting the explicit form of $g_0$, we get:
$$
\begin{array}{c}
\displaystyle{\tau_N =
\sum_{m\geq 0} \frac{1}{m!}
\int_{\CCC ^m} d\mu_1 \ldots d\mu_m
\lvac \psistar_{N-1}(-J_+) \ldots \psistar_0 (-J_+)
\psi_+ (z_1) \ldots \psi_+(z_m)}
\\ \\
\displaystyle{\times \,\, P_-
\psistar_+ (1/\bar z_m)\ldots \psistar_+ (1/\bar z_1)
\psi_0 (-J_-)\ldots \psi_{N-1}(-J_-)  \rvac}.
\end{array}
$$
The next step is to notice that only the term with
$m=N$ contributes to the sum and all other terms vanish.
Indeed, at $m>N$ the state
$$
\psistar_+ (1/\bar z_m)\ldots
\psistar_+ (1/\bar z_1)
\psi_0 (-J_-)\ldots \psi_{N-1}(-J_-)  \rvac
$$
is in fact the null state because the number of annihilation
operators exceeds the number of creation operators while at
$m<N$ the operator
$$
P_- \psistar_+ (1/\bar z_m)\ldots \psistar_+ (1/\bar z_1)
\psi_0 (-J_-)\ldots \psi_{N-1}(-J_-)
$$
is in fact the null operator because $P_-$ multiplied by the
uncompensated positive $\psi$-modes from the right gives $0$
(see (\ref{ferm9a})). Therefore, the expression simplifies to
$$
\begin{array}{c}
\displaystyle{\tau_N =
 \frac{1}{N!}
\int_{\CCC ^N} d\mu_1 \ldots d\mu_N
\lvac \psistar_{N-1}(-J_+) \ldots \psistar_0 (-J_+)
\psi_+ (z_1) \ldots \psi_+(z_N)}
\\ \\
\displaystyle{\times \,\, P_-
\psistar_+ (1/\bar z_N)\ldots \psistar_+ (1/\bar z_1)
\psi_0 (-J_-)\ldots \psi_{N-1}(-J_-)  \rvac}.
\end{array}
$$
Since there are as many annihilation operators
to the right of $P_-$ as creation ones, the state
that they produce from the vacuum is proportional to
the vacuum state itself, i.e.,
$$
\psistar_+ (1/\bar z_N)\ldots
\psistar_+ (1/\bar z_1)
\psi_0 (-J_-)\ldots \psi_{N-1}(-J_-)  \rvac =\rvac C_N,
$$
where the constant $C_N$ is
$$
\begin{array}{lll}
C_N&=&\displaystyle{
\lvac  \psistar_+ (1/\bar z_N)\ldots
\psistar_+ (1/\bar z_1)
\psi_0 (-J_-)\ldots \psi_{N-1}(-J_-)  \rvac}
\\ && \\
&=&\displaystyle{
\det_{1\leq j,k \leq N}\lvac \psistar_{+}(1/\bar z_j)\psi_{k-1} (-J_-)\rvac }.
\end{array}
$$
Because
$$
\begin{array}{c}
\displaystyle{
\lvac \psistar_{+}(1/\bar z_j)\psi_{k-1} (-J_-)\rvac =
\sum_{a,l\geq 0}\bar z_j^l \,  p_a (-t_-) \lvac \psistar_{l}\psi_{k+a-1}\rvac}
\\ \\
\displaystyle{ \phantom{aaaaaaaaaa}=\,
\sum_{a\geq 0} \bar z_j^{k+a-1} p_a (-t_-)\, = \,
\bar z_j^{k-1}e^{-\xi (t_- , \bar z_j)}},
\end{array}
$$
the constant $C_N$ is explicitly given by
\beq\label{CN}
C_N = \Delta_N (\bar z_i)\prod_{l=1}^{N}
e^{-\xi (t_- , \bar z_l)},
\eeq
where we use the convenient short-hand notation for the
Vandermonde determinant:
$$
\Delta_N (z_i) =\det_{1\leq j,k \leq N}
\left ( z_{k}^{j-1}\right )=
\prod_{i>j} (z_i -z_j).
$$
Now, it remains to calculate
$$
\begin{array}{c}
\displaystyle{
\lvac \psistar_{N-1}(-J_+)
\ldots \psistar_0 (-J_+)
\psi_+ (z_1) \ldots \psi_+(z_N)\rvac
=\det_{1\leq j,k\leq N}\lvac \psistar_{j-1}(-J_+)\psi_+(z_k)\rvac}
\\ \\
\displaystyle{
=\, \det_{1\leq j,k\leq N} z_{k}^{j-1}e^{\xi (t_+ , z_k)}\, =\,
\Delta_N (z_i)\prod _{l=1}^{N}
e^{\xi (t_+ , z_l)}},
\end{array}
$$
which can be done in a completely similar manner. Collecting everything
together, we obtain the result:
\beq\label{can4}
\tau_N (t_+, t_-)= \frac{1}{N!}\int_{\CCC ^N}|\Delta_N
(z_i)|^2 \prod_{l=1}^{N}
e^{\xi (t_+ , z_l )-\xi (t_- , \bar z_l)} d\mu (z_l).
\eeq

Assume that $d\mu (z)=e^{-U(z, \bar z)}d^2z$
is a smooth measure on the plane and $t_{-k}=-\bar t_k$, then
the expression $\xi (t_+, z)-\xi (t_- , \bar z)$ is purely real
and the integral (\ref{can4}) has a physical interpretation as
the partition function
of a canonical ensemble of $N$ identical Coulomb particles
in the plane in the background
potential $W(z, \bar z)=-U(z, \bar z)+2{\cal R}e \sum_{k}t_k z^k$:
\beq\label{ZN}
Z_N =\frac{1}{N!}\int_{\CCC ^N}|\Delta_N
(z_i)|^2 \prod_{l=1}^{N}
e^{W(z_l, \bar z_l)}d^2 z_l.
\eeq
It is proportional to the partition function
of the ensemble of normal random $N\times N$ matrices $\Phi$ \cite{ChZ}:
$Z_N \propto \int D\Phi \, e^{\mbox{tr}\, W(\Phi, \Phi^{\dag})}$,
with $z_i$ being their eigenvalues.

If the measure $d\mu$ is concentrated on a curve $\Gamma \subset \CC$,
then the 2D integrals $\int_{\CCC}(\ldots )d^2z$ are reduced to
1D integrals $\int_{\Gamma}(\ldots )|dz|$ along $\Gamma$. This means
that the 2D Coulomb particles are confined to the curve $\Gamma$.
For particular choices of $\Gamma$ the integral (\ref{can4}) yields
the partition functions of random matrix models of certain types
in terms of eigenvalues.
For example, if $\Gamma$ is the real line, one obtains the
partition function of hermitian random matrices and if $\Gamma$ is the
unit circle, the integral (\ref{can4}) becomes identical to (\ref{unitary})
which is the partition function of unitary random matrices.

Let us show that for axially symmetric measures $d\mu$
the operator representation (\ref{can3})
is equivalent to the one suggested by A. Orlov et al
\cite{OSch00}.
For an axially symmetric measure,
the bilinear form in the fermion operators in (\ref{can1}) becomes
diagonal:
$$
\int_{\CCC}\psi_+(z)\psistar _{+}(1/\bar z)
d\mu (z)=\sum_{m,n\geq 0}\psi_n \psistar_m
\int_{\CCC}z^n \bar z^m e^{-U(|z|)}d^2z =
\sum_{n\geq 0} h_n \psi_n \psistar_n,
$$
where
$$
h_n = \int_{\CCC} |z|^{2n}e^{-U(|z|)}d^2z,
$$
(we assume that the measure is smooth with
$U(z, \bar z)=U(|z|)$),
so in this case
\beq\label{can5}
g_0 =\normord \exp \left (\sum_{n\geq 0} (h_n \! -\! 1) \psi_n \psistar_n
\right ) \normord
= \, \exp \left (\sum_{n\geq 0} \log h_n \, \psi_n \psistar_n \right )
\eeq
and the tau-function (\ref{can3}) does have the form
$\lvacN e^{J_+}e^X e^{-J_-}\rvacN$ with
$\displaystyle{X \! = \!\! \sum_{j\in \z}  X_j \normord
\psi_j \psistar_j \normord }$ studied in \cite{OSch00}.
More precisely, it corresponds to a singular
limit of the latter with $X_j =\log h_j$ for $j\geq 0$ and
$X_j \to +\infty$ for all $j<0$. Indeed, writing
$$
e^X = \prod_{j\geq 0}\left ( 1+( e^{X_j}-1)\psi_j \psistar_j\right )\cdot
\prod_{j< 0}\left ( 1+( e^{-X_j}-1)\psistar_j \psi_j \right ),
$$
we see that the first product is equal to
$\prod\limits_{j\geq 0} (1+( h_j -1)\psi_j \psistar_j)=g_0$ while the
limit of the second one is the singular operator $P_+$.

An important example is $U(z, \bar z)=c|z|^2$, then $h_n =\pi c^{-n-1} n!
=\pi c^{-n-1}\Gamma (n+1)$
and $X_n = -(n+1)\log c +
\log \Gamma (n+1)$ (the common constant $\log \pi$ is irrelevant
because the operator $\displaystyle{\sum_{j\in \z} \normord
\psi_j \psistar_j \normord }$ commutes with all elements of the
Clifford algebra). Note that the analytic continuation of this
formula to negative values of $n$ with the help of the
gamma-function automatically implies the required
singular limit
$X_n = +\infty$ at $n<0$. The tau-function (\ref{can4})
for this case has the following expansion in Schur functions
\cite{OSch00,OShi05}:
\beq\label{can4ex}
\tau_N (t_+ , t_-)=\pi^N c^{-N(N+1)/2}
\prod_{k=1}^{N}\Gamma (k) \cdot
\sum_{\lambda}c^{-|\lambda |}(N)_{\lambda}\,
s_{\lambda}(t_+)s_{\lambda}(-t_-).
\eeq
Here $\lambda$ denotes the Young diagram with $\ell (\lambda )$
rows of lengths $\lambda_1 \geq \lambda_2 \geq \ldots
\geq \lambda _{\ell (\lambda )}>0$,
$|\lambda |$ is the total number of boxes in
$\lambda$,
$$
(N)_{\lambda}:=\prod_{i=1}^{\ell (\lambda )}
(N+1 -i)(N+2-i) \ldots (N+\lambda _i -i)
$$
and the sum runs over all Young diagrams
including the empty one (its contribution is $1$).
The Schur functions $s_{\lambda}$ are defined in the
standard way as
determinants of the polynomials $p_k(t)$ (\ref{schur1}):
$$
s_{\lambda}(t)=\det_{1\leq i,j\leq \ell (\lambda )}
p_{\lambda_i -i+j}(t).
$$

The dispersionless limit of the tau-function (\ref{can4})
is achieved via rescaling of times $t_k = T_k /\hbar$, $N= T_0/\hbar$
and tending $\hbar \to 0$. Clearly, this implies
$N\to \infty$. However, in order to get a
meaningful limit the measure $d\mu$ should be chosen appropriately.
In the case of a smooth measure one should set
$d\mu(z)=e^{-\frac{1}{\hbar}U(z, \bar z)}d^2z$, then
\beq\label{lim1}
\tau_N (t_+, t_-)= \exp \left (\frac{F_0}{\hbar^2}+O(\hbar^{-1})\right ),
\quad \hbar \to 0.
\eeq
The function
$F_0=F_0 (\ldots , T_{-1},
T_0 , T_1 , \ldots )$ is what is called
``dispersionless tau-function" \cite{KriW,TakTak}.
When the reality condition $T_{-k}=-\bar T_k$ is imposed,
this function encodes a formal solution
to the inverse potential problem in 2D
and admits a nice geometric/electrostatic description \cite{Z01}.
Set $\sigma (z, \bar z) =\frac{1}{\pi}\p_z \p_{\bar z}U(z, \bar z)$
to be density of the background charges.
We assume that $\sigma >0$. Given a real positive $T_0$
and complex $T_k$'s, $k\geq 1$,
it is a subject of the inverse potential problem to find
a domain ${\sf D}$ in the complex plane such that $T_0$ is
the total charge contained in ${\sf D}$ and $T_k$'s
are harmonic moments of its exterior with respect to the density
$\sigma$:
\beq\label{lim2}
T_0 = \int_{{\sf D}}\sigma (z, \bar z)d^2 z, \quad \quad
T_k = -\frac{1}{k}\int_{\CCC \setminus {\sf D}}\!\!
z^{-k}\sigma (z, \bar z)d^2 z,\quad k\geq 1.
\eeq
Leaving aside the very difficult questions about existence
and uniqueness of the solution, we simply assume, just for
an illustrative purpose, that we are in a situation when
${\sf D}$ is a compact
connected domain (containing the origin). Then $F_0$ is given by
\beq\label{lim3}
F_0 =-\int_{{\sf D}}\! \int_{{\sf D}}
\sigma (z, \bar z) \log \left | z^{-1}-\zeta^{-1}\right |
\sigma (\zeta , \bar \zeta )\, d^2 zd^2\zeta ,
\eeq
which is basically the electrostatic energy of ${\sf D}$ filled
by electric charge with density $\sigma$ (and with a point-like
charge at the origin).
In the case
$U(z, \bar z)=z\bar z$, $\sigma =1/\pi$
one obtains the ``tau-function of analytic curves"
\cite{KKMWZ} which encodes the integrable
structure of parametric families of conformal maps \cite{WZ00}
and the Dirichlet boundary value problem in
$\CC \setminus {\sf D}$ \cite{MWZ02,KMZ05}.
The analytic continuation of the function $F_0$ to general
values of $T_k$ (not necessarily constrained by the condition
$T_{-k}=-\bar T_k$) also has a geometric meaning
in terms of pairs of conformal maps \cite{Teo}.
As the results of \cite{WZ03} suggest, higher terms
in the $\hbar$-expansion of the tau-function (\ref{lim1})
may be related to spectral invariants of the domain ${\sf D}$.

For tau-functions of the form (\ref{can4}) with a singular measure
$d\mu$ concentrated on a contour, the dispersionless
limit does not require introducing any $\hbar$-dependence
of the measure. The leading
$\hbar \to 0$ behavior has the same form (\ref{lim1}) and there is
an integral representation for $F_0$ similar to (\ref{lim3})
with the domain ${\sf D}$ being replaced by a segment of the curve.
However, the parameters $T_k$ in this case do not admit a direct
geometric interpretation.

\section{Partition function of the 2D Coulomb gas as a
tau-function: grand canonical ensemble}

A direct attempt of passing to a grand canonical ensemble via
$\displaystyle{{\cal Z}=\sum_{N\geq 0}e^{\mu N}Z_N}$ with $Z_N$
given by (\ref{ZN}) leads to a quantity whose interpretation as a
tau-function of an integrable hierarchy is not known. A way to
introduce a grand canonical ensemble with good integrable
properties, first suggested by I. Loutsenko et al \cite{LS00},
is to consider the logarithmic gas in the presence
of an ideal conductor, so that each particle interacts not only with
other particles of the gas but also with their ``mirror images" of
opposite charge as well as with its one image. For the reflection to
be globally well-defined the conductor should be a disk.

Let $\DD$ be the unit disk and $\DD ^*$ its exterior. We assume that
particles of the gas with complex coordinates $z_i$
occupy $\DD ^*$ and $\DD$ is an
ideal conductor. Then the mirror images are inside $\DD$
at the points $1/\bar z_i$ and the Coulomb energy of the system is
$$
E_N = \sum_{i< j}^{N} \left (\log |z_i -z_j| +
\log |\bar z_{i}^{-1}-\bar z_{j}^{-1}|\right ) -
\sum_{i,j}^{N} \log |z_i -\bar z_j^{-1}|.
$$
The grand canonical partition function of the system
of identical particles in a
background trapping potential $W$ (which is supposed to include the
chemical potential) is defined as
$$
{\cal Z}=\sum_{N=0}^{\infty}
\frac{1}{N!}
\int_{\DDD ^*}\!\! \ldots \! \int_{\DDD ^*}
e^{E_N +W_N}d^2z_1 \ldots d^2z_N, \quad \quad
W_N = \sum_{l=1}^{N}W(z_l , \bar z_l).
$$
In a more explicit form, it reads
\beq\label{gcan0}
{\cal Z}=\sum_{N=0}^{\infty}
\frac{1}{N!}
\int_{\DDD ^*}\!\! \ldots \! \int_{\DDD ^*}
\prod_{j<k}^{N} \left | \frac{z_j -z_k}{z_j \bar z_k -1}\right |^2
\prod_{l=1}^{N}
\frac{|z_l| e^{W(z_l , \bar z_l)} }{|z_l|^2 \! -\! 1}\, d^2 z_l.
\eeq
In order to ensure convergency of the integrals, the
function $e^{W(z, \bar z)}$
should vanish at $|z|=1$ as
$(|z|^2 -1)^{\alpha}$ with $\alpha >0$.

Let us fix an arbitrary measure $d\mu (z)$ in $\DD ^*$
and consider the following group-like element:
\beq\label{gcan1}
G = \normordbare \exp \left (\int_{\DDD ^*}\psi (z)\psistar (1/\bar z)
d\mu (z)  \right )\normordbare \,.
\eeq
Here we use another normal ordering
$\normordbare (\ldots )\normordbare$, the one with respect
to the completely filled vacuum $\left |-\infty \right >$, which means that
the $\psi$-modes are moved to the left while $\psistar$-modes are moved
to the right.
Expanding the exponent into a series,
one can represent $G$ in a more explicit form:
\beq\label{gcan2}
G= \sum_{N=0}^{\infty} \frac{1}{N!}
\int_{\DDD ^*} \psi (z_1) \ldots \psi (z_N)
\psistar (1/\bar z_N)\ldots \psistar (1/\bar z_1)
d\mu_1 \ldots d\mu_m\,,
\eeq
where $d\mu_j \equiv d\mu (z_j)$.
Using (\ref{ferm5}) it is straightforward to see that
the expectation value
\beq\label{gcan3}
\tau_n^{(G)} (t_+,t_-)=\lvacn e^{J_+}G  e^{-J_-}\rvacn
\eeq
has the structure of partition function of a grand canonical ensemble:
\beq\label{gcan4}
\tau_n^{(G)} (t_+,t_-)=e^{-\sum_{k\geq 1}
kt_k t_{-k}}\sum_{N\geq 0}\frac{1}{N!}
\int_{\DDD ^*}\!\! \ldots \! \int_{\DDD ^*}
\prod_{j<k}^{N} \left | \frac{z_j -z_k}{z_j \bar z_k -1}\right |^2
\prod_{l=1}^{N}e^{\omega (t; z_l, \bar z_l)}
\frac{d\mu (z_l)}{|z_l|^2 \! -\! 1},
\eeq
where
\beq\label{gcan5}
\begin{array}{lll}
\omega (t;z,\bar z) &=& t_0 \log |z|^2 +
\xi (t_+, z)+\xi(t_-, 1/z)-\xi (t_+, 1/\bar z)
-\xi (t_-, \bar z)
\\ && \\
&=&\displaystyle{
t_0 \log |z|^2 +
\sum_{k\geq 1}\Bigl ( t_k (z^k -\bar z^{-k}) - t_{-k}
(\bar z^k -z^{-k})\Bigr )}
\end{array}
\eeq
and $n \equiv t_0$.
Clearly, at $t_{-k}=-\bar t_k$
$\omega (t;z,\bar z)$ is real and can be interpreted as
a harmonic part of the background potential in $\DD ^*$.
More precisely, the identification with (\ref{gcan0}) goes
as follows: set
$W(z, \bar z)=-U(z, \bar z) + \omega (t;z,\bar z)-\log |z|$,
then $d\mu (z)=e^{-U(z, \bar z)}d^2z$.
It is convenient to redefine the tau-function by extracting
the simple factor in front of the sum in (\ref{gcan4}):
$\tilde \tau_n^{(G)} (t_+,t_-)=e^{\sum_{k\geq 1}
kt_k t_{-k}}\tau_n^{(G)} (t_+,t_-)$, then
${\cal Z}=\tilde \tau_n^{(G)} (t_+,t_-)$.

Note that the tau-function $\tilde \tau ^{(G)}$
is formally a $\infty$-soliton tau-function with momenta
of solitons distributed with the measure $d\mu$.
It is the
Fredholm determinant of an integral operator:
\beq\label{gcan6}
\tilde \tau_n^{(G)} (t_+,t_-)
= \det ({\bf 1}+\hat K).
\eeq
Here ${\bf 1}$ is the identity operator and the operator $\hat K$
acts to functions on $\DD ^*$ as follows:
$$
\hat K f(z)=\int_{\DDD ^*}\frac{f(\zeta)
e^{\omega(t;\zeta , \bar \zeta )}}{z\bar \zeta -1}\, d\mu (\zeta).
$$
To see this,
we again calculate the expectation value of (\ref{gcan2})
with the help of (\ref{ferm5}) but now using the
determinant representation of each term. This gives
an expansion of the Fredholm determinant.

As an example, consider the case when
the measure $d\mu$ is concentrated on a circle of
radius $e^{\epsilon}$, $\epsilon >0$. Set $z=e^{\epsilon +i \phi}$,
then $d\mu (z)= e^{\epsilon}d\phi$ and the sum in the r.h.s. of
(\ref{gcan4}) becomes the grand canonical partition function
of the charged particles on the circle $|z|=e^\epsilon$ in the presence of
the ideal conductor:
\beq\label{gcan7}
{\cal Z}= \sum_{N\geq 0}\frac{e^{\mu N}}{N!}
\int_{0}^{2\pi}\!\!\! \ldots \! \int_{0}^{2\pi}
\prod_{j<k}^{N} \left |
\frac{\sin \frac{\phi_j -
\phi_k}{2}}{\sin \frac{\phi_j -\phi_k+2i\epsilon}{2}}\right |^2
\prod_{l=1}^{N}e^{V(\phi_l)}d\phi_l\, ,
\eeq
where the chemical potential $\mu$ is expressed through $t_0$ as
$\mu = 2\epsilon t_0 -\log \left ( e^{\epsilon}- e^{-\epsilon}\right )$
and
$$
V(\phi )=2\sum_{k\in \z}\mbox{sinh}\, (\epsilon k)\, t_k e^{ik\phi}.
$$
This partition function is a compactified version of the one
considered by V. Kazakov et al in \cite{KKN}. As a function
of the ``times'' $t_k$, it is tau-function
of the 2DTL hierarchy.

A slightly more general setting is to take the disk-like
conductor of an arbitrary radius $R$ and to consider the
grand canonical ensemble of charged particles in its exterior.
In fact for any finite $R$ this hardly brings anything new because
the systems at different $R$ can be transformed into
each other by a simple rescaling of $z$ and by redefining
the background potential. However, in the singular limit
$R\to \infty$ a new grand canonical ensemble emerges, which
can be introduced independently without a reference to
any limiting process.
It is defined in the upper half plane $\HH$, with the lower half plane
being an ideal conductor. Its partition function is tau-function
of the KP hierarchy. The limiting procedure is rather sophisticated
and will not be described here. We only remark that it is
similar to the one developed by E. Antonov et al \cite{AO}
for the transition $\mbox{2DTL} \, \to \, \mbox{KP}$ in the
continuum limit of the 2DTL hierarchy.

Let us give the main formulas related to the grand
canonical ensemble in $\HH$ and its operator realization.
Fix an arbitrary measure $d\mu _{\HHH}(z)$ in $\HH$
and consider the following group-like element:
\beq\label{gcan8}
G_{\HHH} = \normordbare \exp \left (i \!
\int_{\HHH} \! \psi (z)\psistar (\bar z)\bar z^{-1}
d\mu _{\HHH} (z)  \right )\normordbare .
\eeq
The expectation value
\beq\label{gcan9}
\tau (t)= \lvac e^{J_+}G_{\HHH}\rvac
\eeq
is KP tau-function depending on the
``times" $t=\{t_1, t_2, \ldots \}$.
(For the abuse of notation we use the same letters as in
(\ref{gcan3}) but one should remember that their meaning
is different, see below.)
A calculation similar to the one for $\DD ^*$ yields
\beq\label{gcan10}
\tau (t)=\sum_{N\geq 0}\frac{1}{N!}
\int_{\HHH}\!\! \ldots \! \int_{\HHH}
\prod_{j<k}^{N} \left | \frac{z_j -z_k}{z_j -\bar z_k}\right |^2
\prod_{l=1}^{N}e^{\xi (t; z_l)- \xi (t; \bar z_l)}\,
\frac{d\mu _{\HHH}(z_l)}{{2\cal I}m \, z_l},
\eeq
which looks like a $\infty$-soliton tau-function with continuously
distributed momenta. (The analogy with soliton solutions of the
KP hierarchy was first noticed in \cite{LS00}.)
In order to ensure convergency on the real line, the measure
should vanish there as $d\mu _{\HHH}(z)\propto ({\cal I}m \, z)^{\alpha}$
with $\alpha >0$.
Assuming that all the $t_k$'s are purely imaginary and
$d\mu _{\HHH}(z)=e^{-U(z, \bar z)}d^2z$,
this tau-function is identical to the partition
function of the grand canonical ensemble of 2D Coulomb particles
in $\HH$ with the background potential $W(z, \bar z) =-U(z, \bar z)
+\xi (t,z)-\xi(t, \bar z)$. Choosing the measure $d\mu_{\HHH}$
concentrated on the line $x+i\epsilon$, one obtains from
(\ref{gcan10}) a non-compact analog of (\ref{gcan7}), which is
precisely the partition function considered in \cite{KKN}.

\begin{figure}[t]
   \centering
        \includegraphics[angle=-00,scale=0.55]{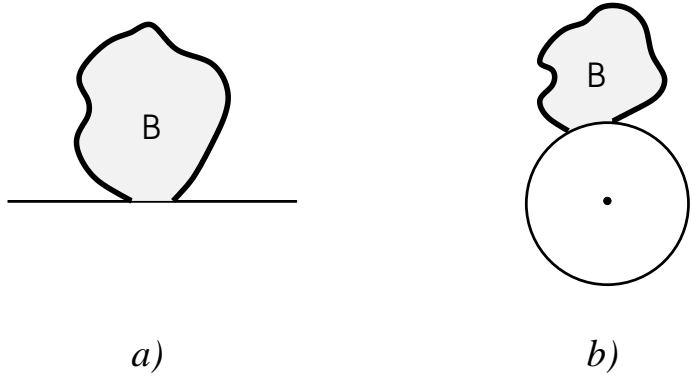}
        \caption{\it  A fat slit ${\sf B}$ a) in the upper half plane,
        b) in the exterior of the unit disk.}
    \label{fig:fs}
\end{figure}

The dispersionless limit of the tau-function (\ref{gcan10})
is again achieved via rescaling of times $t_k = T_k /\hbar$,
setting the measure to be
$d\mu _{\HHH}(z)=e^{-\frac{1}{\hbar}U(z, \bar z)}d^2z$
and tending $\hbar \to 0$. In contrast to the canonical case,
the mean total charge of the system
can not be taken arbitrary but is fixed by the equilibrium
condition. In other words, at $\hbar \to 0$ the leading contribution
to the sum (\ref{gcan10}) comes from its maximal term.
One has:
\beq\label{glim1}
\tau (t)= \exp \left (\frac{\tilde F_0}{\hbar^2}+O(\hbar^{-1})\right ),
\quad \hbar \to 0.
\eeq
The function
$\tilde F_0=\tilde F_0 (T_{1}, T_2 , \ldots )$ is the
tau-function of the dispersionless KP hierarchy \cite{KriW,TakTak}.
When the reality condition ${\cal R}e \, T_{k}=0$ is imposed,
this function admits a nice geometric/electrostatic description
given in \cite{Z09} for the particular case $\sigma =1/\pi$.
Let $\sigma (z, \bar z) =\frac{1}{\pi}\p_z \p_{\bar z}U(z, \bar z)$
be density of the background charges, as before. Given purely imaginary
$T_k$'s, $k\geq 1$, a version of the inverse potential problem
in the upper half plane allows one to find
a domain ${\sf B}$ (a ``fat slit" \cite{Z09}, see
Fig.~\ref{fig:fs}a) in $\HH$ such that
\beq\label{glim2}
T_1 = -2i\, {\cal I}m \int_{{\sf B}}\sigma (z, \bar z)d^2 z, \quad \quad
T_k = \frac{2i}{k}\, {\cal I}m \int_{\HHH \setminus {\sf B}}\!\!
z^{-k}\sigma (z, \bar z)d^2 z,\quad k\geq 2.
\eeq
Then $\tilde F_0$ is given by
\beq\label{glim3}
\tilde F_0 =-\int_{{\sf B}}\! \int_{{\sf B}}
\sigma (z, \bar z) \log \left | \frac{z-\zeta}{z-\bar \zeta}\right |
\sigma (\zeta , \bar \zeta )\, d^2 zd^2\zeta ,
\eeq
which is basically the electrostatic energy of ${\sf B}$ filled
by electric charge with density $\sigma$ in the presence of
an ideal conductor filling the lower half-plane. As a function
of the $T_k$'s it obeys the infinite set of dispersionless Hirota
relations. The corresponding Lax function performs the conformal map
from the upper half plane $\HH$ to the complement of the ``fat slit"
${\sf B}$ in $\HH$.
Increasing $T_1$ and keeping all $T_k$'s with $k\geq 2$ fixed,
one obtains a growth problem of Laplacian type in the upper
half plane, with a specific boundary
condition on the real line, which is associated with the
dispersionless KP hierarchy in the same way as the problem
in the whole plane is associated with the dispersionless 2DTL
hierarchy.

The tau-function (\ref{gcan4}) regarded as a function of
``slow times" $T_k=\hbar t_k$, $k\in \ZZ$,
has a similar $\hbar \to 0$ limit yielding
tau-function of the dispersionless 2DTL hierarchy.
It generates a growth problem for ``fat slits"
in the exterior of the unit disk (Fig.~\ref{fig:fs}b),
which will be discussed elsewhere.

\section*{Acknowledgments}

I thank A. Orlov for illuminating discussions and to
the referee for bringing paper \cite{LS00} to my attention.
This work was supported in part by RFBR grant 08-02-00287,
by RFBR-CNRS grant 09-01-93106-CNRS,
by grant for support of scientific schools NSh-3035.2008.2 and by
Federal Agency for Science and Innovations of Russian Federation
under contract 02.740.11.5029.

\end{document}